\documentclass[prl,superscriptaddress,twocolumn,nobibnotes]{revtex4-2}

\usepackage{graphicx}
\usepackage{amsmath}
\usepackage{amssymb}
\usepackage{amsfonts}
\usepackage{booktabs}
\usepackage{multirow}
\usepackage{url}
\usepackage{tabularx}
\usepackage{array}
\usepackage{siunitx}
\usepackage{xcolor}
\usepackage[colorlinks=true, allcolors=blue]{hyperref}

\newcommand{\paper}{Letter}
\newcommand{\Unu}{\ensuremath{\mathrm{UHE}\nu}}

\newcommand{\atUC}{\affiliation{Dept.~of Physics, Dept.~of Astronomy and Astrophysics, Enrico Fermi Institute, Kavli Institute for Cosmological Physics, University of Chicago, Chicago, IL 60637}}
\newcommand{\atKU}{\affiliation{Dept.~of Physics and Astronomy, University of Kansas, Lawrence, KS 66045}}
\newcommand{\atOSU}{\affiliation{Dept.~of Physics, Center for Cosmology and AstroParticle Physics, The Ohio State University, Columbus, OH 43210}}
\newcommand{\atUW}{\affiliation{Dept.~of Physics, Wisconsin IceCube Particle Astrophysics Center, University of Wisconsin-Madison, Madison,  WI 53706}}
\newcommand{\atNTU}{\affiliation{Dept.~of Physics, Graduate Institute of Astrophysics, Leung Center for Cosmology and Particle Astrophysics, National Taiwan University, Taipei, Taiwan}}
\newcommand{\atULB}{\affiliation{Universit\'{e} Libre de Bruxelles, Science Faculty CP230, B-1050 Brussels, Belgium}}
\newcommand{\atUMD}{\affiliation{Dept.~of Physics, University of Maryland, College Park, MD 20742}}
\newcommand{\atPSUigc}{\affiliation{Center for Multi-Messenger Astrophysics, Institute for Gravitation and the Cosmos, Pennsylvania State University, University Park, PA 16802}}
\newcommand{\atPSUphys}{\affiliation{Dept.~of Physics, Pennsylvania State University, University Park, PA 16802}}
\newcommand{\atPSUast}{\affiliation{Dept.~of Astronomy and Astrophysics, Pennsylvania State University, University Park, PA 16802}}
\newcommand{\atVUB}{\affiliation{Vrije Universiteit Brussel, Brussels, Belgium}}
\newcommand{\atUNL}{\affiliation{Dept.~of Physics and Astronomy, University of Nebraska, Lincoln, Nebraska 68588}}
\newcommand{\atWhittier}{\affiliation{Dept.~Physics and Astronomy, Whittier College, Whittier, CA 90602}}
\newcommand{\atNUU}{\affiliation{Dept.~of Energy Engineering, National United University, Miaoli, Taiwan}}
\newcommand{\atChiba}{\affiliation{International Center for Hadron Astrophysics, Chiba University, Chiba 263-8522 Japan}}
\newcommand{\atNPU}{\affiliation{Dept.~of Applied Physics, National Pingtung University, Pingtung City, Pingtung County 900393, Taiwan}}
\newcommand{\atUCL}{\affiliation{Dept.~of Physics and Astronomy, University College London, London, United Kingdom}}
\newcommand{\atUD}{\affiliation{Dept.~of Physics, University of Delaware, Newark, DE 19716}}
\newcommand{\atUNAF}{\affiliation{Unaffiliated, Madison, WI 53706}}
\newcommand{\atNDL}{\affiliation{National Nano Device Laboratories, Hsinchu 300, Taiwan}}
\newcommand{\atUtah}{\affiliation{Dept.~of Physics and Astronomy, University of Utah, Salt Lake City, UT 84112}}
\newcommand{\atDESY}{\affiliation{Deutsches Elektronen-Synchrotron DESY, Platanenallee 6, 15738 Zeuthen, Germany}}

\begin{document}



\title{New Constraints on the Ultrahigh Energy Neutrino Flux\\from the First Array-Wide Search with the Askaryan Radio Array}


 \author{N.~Alden}\atUC
 \author{S.~Ali}\atKU
 \author{P.~Allison}\atOSU
 \author{D.Z.~Besson}\atKU
 \author{A.~Bishop}\email{abigail.bishop@icecube.wisc.edu}\atUW
 \author{P.~Chen}\atNTU
 \author{Y.C.~Chen}\atNTU
 \author{Y.-C.~Chen}\atNTU
 \author{S.~Chiche}\atULB
 \author{B.A.~Clark}\email{baclark@umd.edu}\atUMD
 \author{A.~Connolly}\atOSU
 \author{K.~Couberly}\atKU
 \author{A.~Cummings}\atPSUigc\atPSUphys\atPSUast
 \author{P.~Dasgupta}\email{dasgupta.80@osu.edu}\atOSU
 \author{K.D.~de~Vries}\atVUB
 \author{C.~Deaconu}\atUC
 \author{M.A.~DuVernois}\atUW
 \author{J.~Flaherty}\atOSU
 \author{P.~Giri}\email{pgiri4@huskers.unl.edu}\atUNL
 \author{J.~Hanson}\atWhittier
 \author{K.D.~Hoffman}\atUMD
 \author{M.-H.~Huang}\atNTU\atNUU
 \author{K.~Hughes}\atOSU
 \author{A.~Ishihara}\atChiba
 \author{A.~Karle}\atUW
 \author{J.L.~Kelley}\atUW
 \author{M.-C.~Kim}\atChiba
 \author{I.~Kravchenko}\atUNL
 \author{R.~Krebs}\atPSUigc\atPSUphys
 \author{C.Y.~Kuo}\atNTU
 \author{U.A.~Latif}\atVUB
 \author{C.H.~Liu}\atUNL
 \author{T.C.~Liu}\atNTU\atNPU
 \author{W.~Luszczak}\atOSU
 \author{A.~Machtay}\email{machtay.1@osu.edu}\atOSU
 \author{M.S.~Muzio}\email{muzio@wisc.edu}\atUW
 \author{J.~Nam}\atNTU
 \author{R.J.~Nichol}\atUCL
 \author{A.~Novikov}\atUD
 \author{A.~Nozdrina}\atOSU
 \author{E.~Oberla}\atUC
 \author{C.W.~Pai}\atNTU
 \author{Y.~Pan}\atUD
 \author{C.~Pfendner}\atUNAF
 \author{N.~Punsuebsay}\atUD
 \author{J.~Roth}\atUD
 \author{A.~Salcedo-Gomez}\email{salcedogomez.1@osu.edu}\atOSU
 \author{D.~Seckel}\atUD
 \author{M.F.H.~Seikh}\email{fulhossain@ku.edu}\atKU
 \author{Y.-S.~Shiao}\atNTU\atNDL
 \author{S.C.~Su}\atNTU
 \author{S.~Toscano}\atULB
 \author{J.~Torres}\atOSU\atUtah
 \author{N.~van~Eijndhoven}\atVUB
 \author{A.~Vieregg}\atUC
 \author{M.~Vilarino~Fostier}\atULB
 \author{M.-Z.~Wang}\atNTU
 \author{S.-H.~Wang}\atNTU
 \author{P.~Windischhofer}\atUC\atDESY
 \author{S.A.~Wissel}\atPSUigc\atPSUphys\atPSUast
 \author{S.~Yoshida}\atChiba
\collaboration{ARA Collaboration}\noaffiliation

\date{\today}

\begin{abstract}
Ultrahigh energy neutrinos ($E_\nu \gtrsim 100$\,PeV) are detectable through Askaryan emission---the nanosecond-scale radio impulse emitted by relativistic particle cascades in a dielectric medium. The Askaryan Radio Array (ARA), an Askaryan detector array located at the South Pole, has been taking data for more than a decade. We present the first array-wide search for diffuse ultrahigh energy neutrinos with ARA. The search found no events on an expected background of $0.13\pm0.01\text{ (stat.)}^{+0.05}_{-0.06}\text{ (syst.)}$~events. The resulting upper limit is the most stringent from $50$\,EeV to $1000$\,EeV and provides the first constraints on the flux reported by the KM3NeT Collaboration that are complementary to optical Cherenkov and particle detectors.
This limit also constrains UHECR source models, and excludes some models with trans-GZK proton populations at the $90\%$~confidence level.
This represents the largest Askaryan detector dataset ever analyzed, demonstrating the feasibility and sensitivity of searches with large detector arrays acting as a single observatory.
It also underscores the high acceptance and exposure of Askaryan detector arrays to ultrahigh energy neutrinos.
\end{abstract}


\maketitle

\textbf{\textit{Introduction ---}} How and where nature accelerates ultrahigh energy cosmic rays (UHECRs, $E_\mathrm{CR} \gtrsim 10^{18}$\,eV) is a longstanding open question in physics.
Understanding the origins of ultrahigh energy cosmic rays would unlock opportunities to probe extreme astrophysical environments and particle physics beyond LHC energies~\cite{Coleman:2022abf}.
While substantial progress has been made in understanding the broad properties of cosmic rays (that they are extragalactic in origin with a mixed chemical composition)~\cite{Roth:2025hqe,Kido:2025mey}, progress towards directly identifying their sources continues to be hindered by observational challenges, particularly their short energy loss length and deflections in the Galactic magnetic field~\cite{Unger:2023lob,PierreAuger:2024hlp,Bister:2024ocm}.
However, ultrahigh energy neutrinos (\Unu{}, $E_\nu \gtrsim 10^{17}$\,eV) produced through cosmic ray interactions provide a pathway to overcome these challenges, since they are essentially unattenuated and do not experience deflections in propagation. Cosmic rays produce neutrinos through photohadronic and hadronic interactions inside their sources and in extragalactic propagation~\cite{Ackermann:2022rqc}. For this reason, ultrahigh energy neutrinos can be used to probe the properties of, and potentially identify, ultrahigh energy cosmic ray sources.

\par
The expected flux of ultrahigh energy neutrinos is unknown, with theoretical predictions spanning several orders of magnitude, but is strongly motivated by cosmic ray data and flux models~\cite{Ackermann:2022rqc,Muzio:2025xen}. Intriguingly, the KM3NeT Collaboration recently observed a $220^{+570}_{-110}$\,PeV~\cite{KM3NeT:2025npi} neutrino candidate.
However, attempts by larger detectors to measure the flux have only resulted in upper limits~\cite{IceCubeCollaborationSS:2025jbi,ANITA:2019wyx} indicating the flux is quite low, at the level of $E_\nu^2\Phi_\nu \lesssim 10^{-8}$\,GeV/cm$^2$/s/sr at $1$\,EeV (corresponding to roughly $1$\,neutrino/yr/$10$\,km$^3$ of detection volume), and is expected to fall with energy. For this reason, much larger detectors are required to measure the flux at $\gtrsim1$\,EeV energies than are practical with optical Cherenkov methods.
Askaryan emission~\cite{Askaryan:1961pfb,ANITA:2006nif}, the nanosecond-scale radio impulse generated by relativistic particle cascades in a dielectric medium, provides a method for neutrino detection which is scalable to the $\gtrsim 10^{2}$\,km$^3$ volumes required to further probe the ultrahigh energy neutrino flux. 

\par
\begin{figure}
    \centering
    \includegraphics[width=\linewidth]{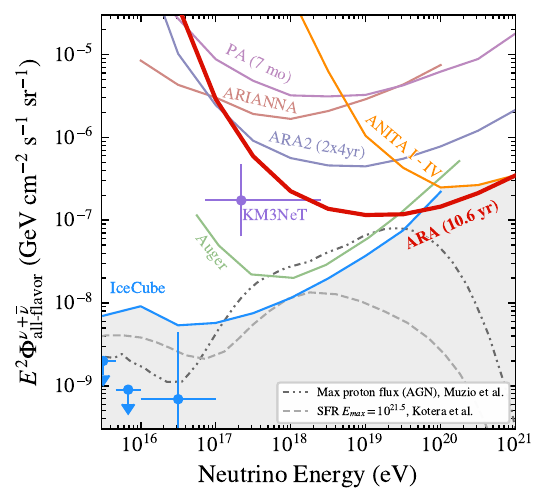}
    \caption{
        The $90\%$ CL upper limit on the all-flavor diffuse UHE neutrino flux set by this analysis (thick red line). Previous constraints and measurements from IceCube~\cite{IceCubeCollaborationSS:2025jbi, IceCube:2025ewu}, KM3NeT~\cite{KM3NeT:2025npi}, Auger~\cite{PierreAuger:2023pjg}, ANITA I--IV~\cite{ANITA:2019wyx}, ARIANNA~\cite{Anker:2019rzo}, and earlier ARA analyses~\cite{ARA:2019wcf,ARA:2022rwq} are also shown (colored lines). The optimization was performed on a flux model corresponding to the global upper limit (boundary of grey region). All upper limits are converted to equivalent all-flavor limits using decade-wide bins. Adjustments are not made for differences in the assumed spectral index within each bin.  
        Illustrative UHE neutrino flux models~\cite{Kotera:2010yn,Muzio:2023skc} are presented (grey dashed lines). 
        \label{fig:limit}
    }
\end{figure}

Located near the South Pole, the Askaryan Radio Array (ARA) leverages this detection principle through an array of stations, each composed of radio antennas deployed up to $200$\,m deep in the ice.
ARA consists of five stations and has been taking data since the deployment of a prototype instrument in 2011~\cite{ARA:2011testbed}. Recently, ARA reported the first in-ice observation of Askaryan radiation generated by impacting cosmic ray showers~\cite{ARA:2025qit}.
In this \paper{}, we present the first array-wide search for diffuse ultrahigh energy neutrinos with ARA. We observe no events, and set the limit shown in Fig.~\ref{fig:limit}, which provides the strongest constraints from $50$\,EeV to $1000$\,EeV.
This search demonstrates the ability for large Askaryan detector arrays to perform a sensitive search using an inhomogeneous array of detectors as a single observatory, while controlling backgrounds, even near anthropogenic sources.
A more extensive discussion of this search's methods can be found in a companion paper~\cite{prd}.

\begin{figure}[h!]
    \centering
    \includegraphics[width=\linewidth]{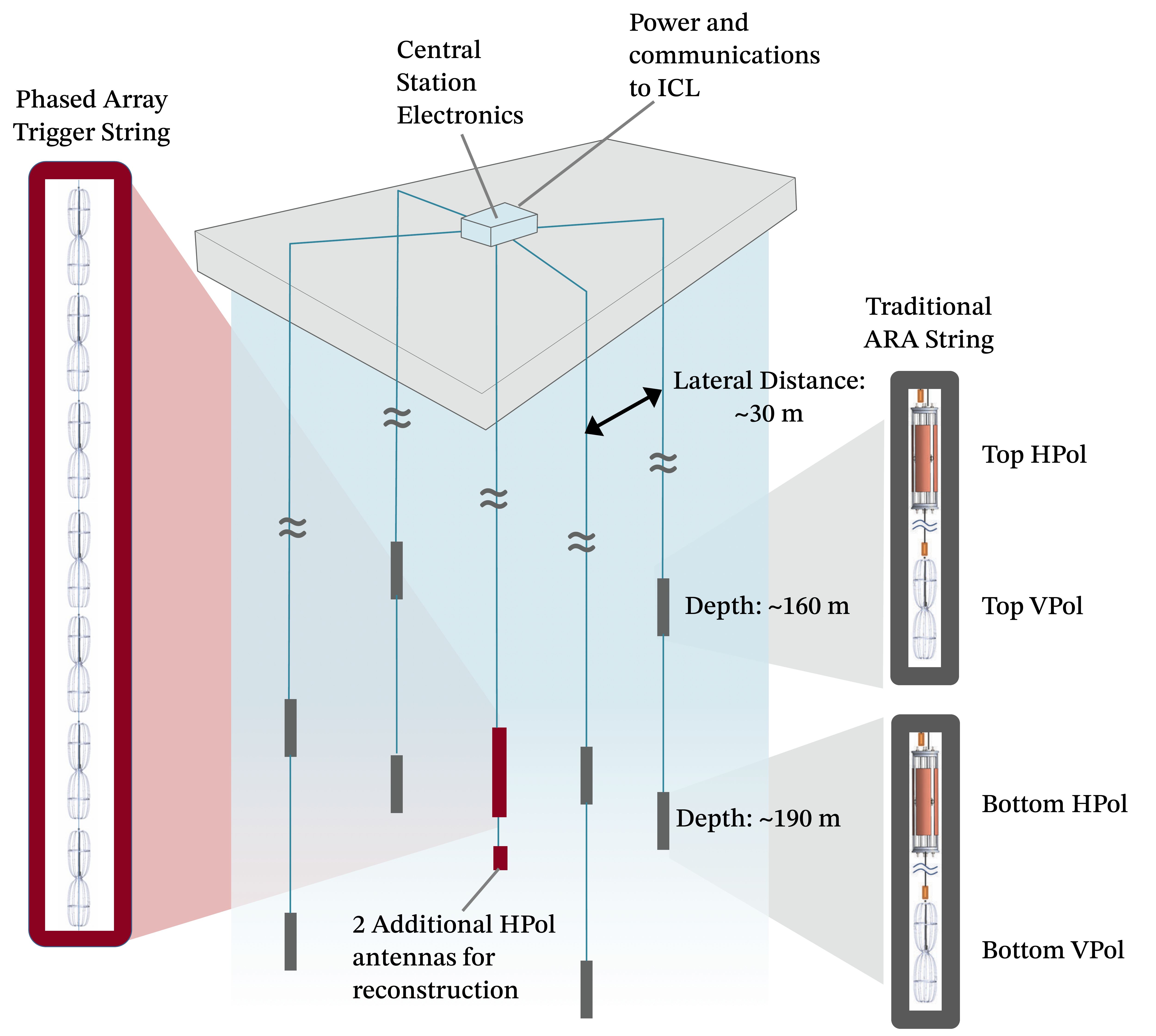}
    \caption{
        ARA's station layout, shown for station A5. Each ARA station has four holes drilled into the ice and instrumented with radio antennas (shown in grey). A5 additionally has the phased array (PA) detector string (shown in red) deployed in the center of the station.
        \label{fig:ARA_station}
    }         
\end{figure}

\textbf{\textit{Detector ---}} The ARA detector has five stations, A1--5, each composed of four strings of four antennas (see Fig.~\ref{fig:ARA_station}) with broadband $150-850$\,MHz sensitivity. Each string is deployed up to $200$\,m depth, except for A1 whose strings are deployed to $100$\,m depth.
The stations are roughly 2\,km apart, and are laid out on a hexagonal grid.
Strings host two antenna pairs that are separated by $\sim30$\,m to form a roughly cubical lattice. Each pair has one antenna sensitive to vertically-polarized signals (VPols) and one sensitive to horizontally-polarized signals (HPols). A station triggers when three like-polarization antennas observe excess power within a coincidence window. The threshold on each station is dynamically adjusted to maintain a trigger rate of not more than $5$\,events/s. Additionally, each station has pulsers for calibration and to monitor detector performance.

\par
The fifth station of ARA, A5, hosts an additional subdetector called the phased array (PA)~\cite{ARA:2018phased}. The PA consists of a single string at the center of the ``standard" A5 detector.
This string hosts seven VPols and two HPols with a close $1-2$\,m spacing.
This compact design enables the PA to use an interferometric trigger~\cite{Vieregg:2015baa}, which significantly decreases the signal-to-noise ratio (SNR) trigger threshold, making the PA more sensitive to low-amplitude signals~\cite{ARA:2018phased}.
The PA threshold is dynamically tuned to maintain a trigger rate of not more than $11$\,events/s.

\par
We consider data taken from January 2013 through December 2023 across all five stations and the PA in this search, corresponding to $28$\,station-years of livetime.
Stations were added to the array over the time period considered in this search:
A1 in 2012, A2/3 in 2013, and A4/5 and PA in 2018.
The stations themselves also had distinct periods of operation (e.g.\ due to differences in trigger windows), which we call ``configurations"~\cite{prd}.
As a result of this evolution, the sensitivity of the array changed over this period.
We consider $50$~distinct configurations of the full array when calculating its overall sensitivity~\cite{ARA:2026wjc}. In view of the proximity of A5 and PA, we only consider data from the more sensitive PA when both detectors were running. Neutrino simulations were performed for each of these array-wide configurations using \texttt{NuLeptonSim}~\cite{Cummings:2023iuw} and \texttt{AraSim}~\cite{ARA:2026wjc} assuming an $E^{-1}$ spectrum (see~\cite{prd} for more details). These array-wide simulations accurately capture neutrinos triggering multiple stations and the time-dependent sensitivity of the detector.

\textbf{\textit{Event Selection ---}} To isolate diffuse \Unu{} in our dataset, we develop an event selection consisting of a series of quality cuts to establish an analysis-quality dataset, and background cuts to remove non-neutrino signals.
To mitigate biases, the event selection was developed on a randomly selected $10\%$ sample, while the remaining $90\%$ was kept ``blinded" until the selection and background estimates were finalized~\cite{Klein:2005di}.
We first prepare the data via a processing pipeline that calibrates and interpolates waveforms before removing anthropogenic continuous wave contamination~\cite{ANITA:2018impulsivity,ARA:2025scq}, applying a frequency-dependent phase shift to correct for the detector response, and bandpassing.

After preparing the data, we calculate a series of analysis variables, such as the SNR, impulsivity, and the reconstructed emission direction (calculated via an interferometric reconstruction on a grid of emission points~\cite{prd}), that are used to distinguish neutrino signals from background events. 
Qualitatively, diffuse neutrinos are expected to produce a fast $\mathcal{O}(\textrm{ns})$, impulsive signal correlated across antennas that originates in deep ice, and is isolated in time.
The most abundant background is thermal blackbody emission from the ice, which is comparatively weaker in amplitude and uncorrelated across antennas.
The most difficult backgrounds to mitigate are those from near and above the surface of the ice, including ``anthropogenics" (e.g.\ snowmobiles and airplanes) and cosmic rays producing radio emission.

\par
Four data quality cuts were applied to distill an analysis-quality dataset. First, a pre-filter is used to remove thermal-like events from the processing chain. This pre-filter was developed on the $10\%$~data after isolating thermal-like events, and is only used during processing of the $100\%$~dataset to reduce computational overhead. Second, we applied livetime cuts to remove data-taking runs which had significant anthropogenic contamination or took place during known calibration campaigns.
Additionally, the first minute of each data-taking run (typically $3$\,h) was removed from the analysis, since the station's trigger thresholds are stabilizing in this period. Third, cuts are applied to remove events that are consistent with emission from the detector itself.
This manifests as an event in which all of the channels on a single string exhibit a large power deposition.
This class of events was observed in previous ARA searches~\cite{ARA:2019wcf} and is hypothesized to be due to electrostatic discharge occurring near the antennas.
Finally, due to its central role in background removal, we apply a reconstruction quality cut that requires events to have a well-localized interferometric reconstruction of their emission direction.
After these quality cuts, $25.9$\,station-years remain,
corresponding to $10.6$\,years of array-wide livetime.

\par
Backgrounds from anthropogenic sources, cosmic rays, and thermal backgrounds are removed from the dataset using a series of five cuts, which we describe at a high-level. Detailed information about these cuts, their efficiencies, and their effectiveness at background removal can be found in~\cite{prd}.
First, we apply a cut to remove events whose emission point reconstructs to directions (typically within about $8^\circ$) of the local calibration pulser.
For A1, a similar cut is applied in the direction of the South Pole due to its proximity to anthropogenic sources and its shallow depth. 
Second, a spatiotemporal cut is applied to remove backgrounds that cannot be confined to a single direction or that are short-lived, such as airplanes and snowmobiles.
Third, two cuts are applied to remove any remaining backgrounds from the direction of the surface, such as anthropogenic backgrounds and cosmic rays.
These surface cuts are constructed to mitigate mis-reconstructions of surface backgrounds to non-surface directions, and have thresholds between roughly $-7^\circ$ and $20^\circ$ elevation.
Fourth, a cut is applied to remove thermal-like backgrounds from the dataset.
The latter two cuts will be discussed in further detail below.
Each of the previous cuts is applied at the station-level.
The fifth and final cut is a time isolation cut, which is applied array-wide.
The time isolation cut removes any remnant events that occur within $5.8$\,h of each other~\cite{prd}, based on the expectation that diffuse neutrino events should arrive isolated from each other in time.

\begin{figure}
    \centering
    \includegraphics[width=\linewidth]{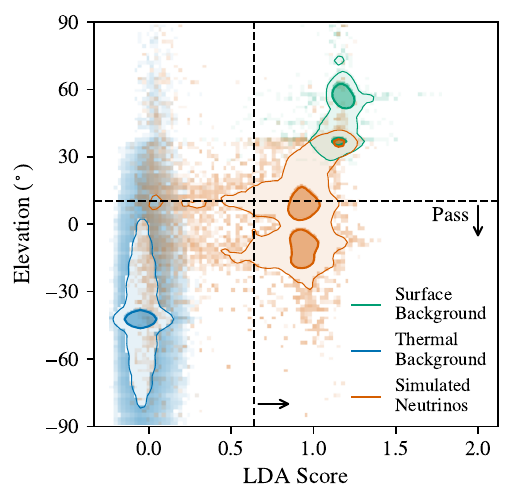}
    \caption{
        Distribution of simulated neutrino (orange), surface background (green), and thermal background (blue) populations compared to the elevation and LDA score thresholds (black dashed lines) for A3 in array-wide configuration 3.
        Contours (solid lines) indicate constant-density boundaries corresponding to the $50\%$ and $90\%$~containment regions.
        The signal region and cut regions are indicated.
        The tail of the surface background population is parametrized as a KDE in elevation, while the tail of the thermal background population is parametrized as an exponential in LDA score. The long tails of these populations force the optimizer to set strong cuts even while the bulk is compact.
        We note that the distribution of thermal events in elevation is biased due to known instrumental artifacts that arise when correlating low SNR events~\cite{clark2019, lu2020}.
        \label{fig:thermal_surface_plane}
    }
\end{figure}

\par
The surface background and thermal-like background cuts removed the most background events: $15\%$ and $100\%$ of events remaining after earlier cuts, respectively. However, they were also the least efficient in terms of the number of simulated neutrino events that they retain in the event selection: $56\%$ and $85\%$ array-wide for an $E^{-1}$ spectrum, respectively~\cite{prd}.
Two strategies are used to mitigate this efficiency loss.
First, we use a linear discriminant analysis (LDA) to form a single score from multiple analysis variables that maximizes the separation of the thermal-like background population from the simulated neutrino population.
This minimizes the overlap between these populations, optimizing the cut's overall efficiency.
An LDA is trained for each configuration of a station.
The tail of the thermal-like background population is parametrized in terms of this LDA score via an exponential model.
Next, we determine the thresholds for the thermal-like background and surface background cuts through an optimization for the search's $5\sigma$~discovery potential of a benchmark flux equivalent to the current global upper limit (UL)~\cite{IceCubeCollaborationSS:2025jbi,ANITA:2019wyx}, which is visualized as the boundary of the grey region in Fig.~\ref{fig:limit}.
To do this, we build a surface background dataset and model this population using a Gaussian kernel density estimator (KDE).
We optimize array-wide over the full livetime of the array to set $144$~thermal-like cut thresholds ($1$ per active station per array-wide configuration) and $6$~surface thresholds ($1$ per standard station and the PA).
This optimization balances maximizing the array-wide efficiency, which must take into account events triggering multiple stations, with minimizing the leakage of the thermal-like and surface background populations into the signal region.
The resulting thresholds for a single station in an example array-wide configuration are shown in Fig.~\ref{fig:thermal_surface_plane}, compared to thermal-like \& surface background populations and the simulated neutrino population.
As can be seen, the neutrino population separates from the thermal-like background in LDA score, and separates from the surface-like background in zenith.
On the benchmark flux, the overall efficiency is $28\%$, with the largest loss coming from the surface-rejection cuts. These uniquely reject about $30\%$ of neutrino events array-wide for an $E^{-1}$ spectrum.

\textbf{\textit{Results ---}} A single event, observed by the PA, passed the event selection before application of the time isolation cut.
Upon investigation, this event was determined to be a background, believed to originate from electrostatic discharge on an A5 string~\cite{prd}, and was removed via an \textit{a posteriori} cut. This removed an additional $21.9$\,s of livetime from the dataset.
Therefore, the search resulted in $0$~passing events on an expected background of $0.13\pm0.01\text{ (stat.)}^{+0.05}_{-0.06}\text{ (syst.)}$~events. A detailed discussion of the calculation of these uncertainties can be found in~\cite{prd}.
Given that no candidate events were observed, we place a $90\%$~confidence level (CL) upper limit on the diffuse flux of UHE neutrinos, using the Feldman--Cousins (FC) construction~\cite{Feldman:1997qc} with the correction from~\cite{Roe:1998zi}, shown in Fig.~\ref{fig:limit}.
This limit is the strongest across all experimental techniques from $50$\,EeV to $1000$\,EeV. 

\par 
This result demonstrates the ability of searches to successfully control backgrounds array-wide, even in the presence of nearby anthropogenic sources. The overall background expectation has three contributions: untagged calibration pulsers, thermal noise, and non-thermal-like events from the surface region. The thermal and surface contributions dominate the expected background at roughly comparable levels. 

\par
The overall exposure of this search is subject to systematic uncertainties from the ice attenuation and index of refraction models, as well as, models of Askaryan emission and the neutrino-nucleon cross section.
These are discussed and quantified at length in the companion paper~\cite{prd}.
At low energies, uncertainties in the Askaryan model dominate ($\sim50\%$), while at high energies the uncertainties in the cross section dominate ($\sim150\%$). 

\textbf{\textit{Discussion ---}} The new upper limit established in this \paper{} is the most stringent from $50$\,EeV to $1000$\,EeV, providing new constraints on the composition and sources of UHECRs. Moreover, this search provides new and complementary probes of UHE neutrino fluxes in the literature. Table~\ref{tab:exclusions} summarizes the expected number of events at analysis-level, $\mu$, for several \Unu{} flux models, as well as the level at which they are constrained by this analysis. We quantify this exclusion in two ways: the Poisson rejection significance (``Significance'') calculated from the $p$-value, $p=\mathrm{Poiss}(n=0~\vert~\mu)$; and, the confidence level, $\alpha$, at which the model can be excluded under the Feldman--Cousins construction. The latter is calculated by solving $\mathrm{FC}_\alpha(n,b) = \mu$, where $\mathrm{FC}_\alpha$ is the Feldman--Cousins upper limit at $\alpha\%$ CL, $n=0$ is the observed number of events, and $b=0.13$ is the background expectation.

In particular, this analysis constrains UHECR source models with trans-GZK components (i.e.\ that have a maximum energy-per-nucleon beyond the GZK threshold of $10^{19.7}$\,eV). These models are typically proton-dominated and so have been strongly constrained by UHECR composition data~\cite{Tkachenko:2025sxm}. Despite this, they are of significant interest since their sources would be among the most extreme accelerators in the universe and the most readily identifiable, due to the minimal deflections of their CRs and sizable \Unu{} flux. This result excludes the Auger composition-compatible trans-GZK UHECR source model of~\cite{Muzio:2023skc} at roughly the $90\%$~CL, as it predicts $2.43$~events to be observed at the analysis-level.
This exclusion demonstrates the additional (source evolution-dependent) constraints this result places on the proton fraction above $30$\,EeV produced by UHECR sources~\cite{Muzio:2023skc}, complementing those from UHECR observatories. 

The search also probes the energy range of the $220$\,PeV neutrino candidate reported by the KM3NeT Collaboration~\cite{KM3NeT:2025npi}. Assuming an $E^{-2}$~spectrum over the energy range encompassed by the event uncertainties, this analysis would expect $1.6$~detected events, which results in a $1.6\sigma$ tension given the KM3NeT observation and non-observation of this search. The absence of observed candidates therefore begins to constrain fluxes of this magnitude with a technique complementary to optical Cherenkov detectors.

Given that KM3-230213A was in ARA's field of view, future targeted point source searches could further constrain transient scenarios and develop the complementarity of radio Askaryan detection to optical Cherenkov detection. 

\begin{table}
    \centering
    \begin{tabularx}{\columnwidth}{>{\centering\arraybackslash}X|S[table-format=1.1]|S[table-format=-1.2,table-space-text-post={\,\(\sigma\)},table-align-text-post=false]|S[table-format=2.2,table-space-text-pre={\(<\)},table-align-text-pre=true,table-space-text-post={\,\%},table-align-text-post=false]}
    \toprule
    \textbf{Flux Model} & {\textbf{Analysis}} & {\textbf{Significance}} & {\textbf{FC CL}} \\
    \midrule
    Global UL \cite{IceCubeCollaborationSS:2025jbi, ANITA:2019wyx} & 5.3 & 2.6{\,\(\sigma\)} & 99{\,\%} \\
    Muzio~\cite{Muzio:2023skc} & 2.4 & 1.3{\,\(\sigma\)} & 90{\,\%} \\
    KM3NeT~\cite{KM3NeT:2025npi} & 1.6 & 0.8{\,\(\sigma\)} & 77{\,\%} \\
    Kotera~\cite{Kotera:2010yn} & 0.4 & -0.4{\,\(\sigma\)} & 27{\,\%} \\
    100\% Protons~\cite{Ahlers:2012rz} & 0.2 & -0.9{\,\(\sigma\)} & {\(<\)}1{\,\%} \\
    Fang Pulsars~\cite{Fang:2013vla} & 0.2 & -0.9{\,\(\sigma\)} & {\(<\)}1{\,\%} \\
    \bottomrule
    \end{tabularx}
    \caption{\label{tab:exclusions}
        The analysis-level event expectation for various models, along with the exclusion level, given that this analysis resulted in no neutrino candidates. We quantify the exclusion level in two ways: the Poisson rejection significance and the Feldman--Cousins confidence level at which the model is excluded. 
    }
\end{table}

\textbf{\textit{Conclusion ---}} In this \paper{} we present the first array-wide, full livetime search for diffuse \Unu{} with ARA. The search resulted in $0$~observed events on an expected background of $0.13\pm0.01\text{ (stat.)}^{+0.05}_{-0.06}\text{ (syst.)}$~events.
The resulting upper limit is the most constraining limit overall from $50$\,EeV to $1000$\,EeV.
This limit places constraints on UHECR source models, complementing those from UHECR observatories. In particular, models with trans-GZK proton populations are constrained, such as~\cite{Muzio:2023skc}, which is excluded at $90\%$~CL. 
ARA will continue to take data, with efforts ongoing to develop next-generation data acquisition systems (``ARA-Next") that will have yet lower trigger thresholds~\cite{ARA:2024erb} and add additional livetime.

This search has shown that a distributed radio array can be analyzed as a single observatory and the backgrounds can be controlled, despite proximity to anthropogenic sources. This is important for the proposed IceCube-Gen2 Radio~\cite{IceCube-Gen2:2020qha,IceCubeGen2_TDR} neutrino observatory, which will be located in the same environment as ARA. Further, this search demonstrates that Askaryan detector arrays have come to maturity and are beginning to provide world-leading sensitivity to the UHE neutrino flux.
The analysis presented here provides a foundation to build upon for future searches on ARA and next-generation detectors, like RNO-G~\cite{RNO-G:2020rmc} (under construction) and IceCube-Gen2 Radio. 

\begin{acknowledgments}
\textbf{\textit{Acknowledgments ---}} This analysis was conducted by A.~Bishop, B.~Clark, P.~Dasgupta, P.~Giri, A.~Machtay, M.~Muzio, A.~Salcedo-Gomez, and M.~F.~H.~Seikh.
\noindent
The ARA Collaboration is grateful for support from the National Science Foundation through Award No.~2013134 and~2310095.
The ARA Collaboration designed, constructed, and now operates the ARA detectors. 
We would like to thank IceCube, and specifically the winterovers, for the support in operating the detector. 
Data processing and calibration, Monte Carlo simulations of the detector and of theoretical models, and data analyses were performed by a large number
of collaboration members, who also discussed and approved the scientific results presented here. 
We are thankful to Antarctic Support Contractor staff, a Leidos unit for field support and enabling our work on the harshest continent. 
We thank the National Science Foundation (NSF) Office of Polar Programs and Physics Division for funding support.
We further thank the Taiwan National Science Council's Vanguard Program NSC 92-2628-M-002-09 and the Belgian F.R.S.-FNRS and FWO.
K.~Hughes thanks the NSF for support through the Graduate Research Fellowship Program Award No.~1746045. 
A.~Connolly thanks the NSF for Awards No.~1806923 and No.~2209588 and also acknowledges the Ohio Supercomputer Center. A.~Connolly and K.~Hughes would like to also thank the Center for Cosmology and AstroParticle Physics at the Ohio State University for funding, travel, and other support.
S.~A.~Wissel thanks the NSF for support through CAREER Award No.~2033500.
A.~Vieregg, C.~Deaconu, N.~Alden, and P.~Windischhofer thank the NSF for Award No.~2411662 and the Research Computing Center at the University of Chicago
for computing resources.
R.~Nichol thanks the Leverhulme Trust for their support. 
K.D.~de~Vries is supported by European Research Council under the European Union's Horizon research and innovation program (Grant Agreement 763 No.~805486). 
D.~Besson, I.~Kravchenko, and D.~Seckel thank the NSF for support through the IceCube EPSCoR Initiative (Award ID No.~2019597). 
M.S.~Muzio thanks the NSF for support through the MPS-Ascend Postdoctoral Fellowship under Award No.~2138121. 
A.~Bishop thanks the Belgian American Education Foundation for their Graduate Fellowship support.

\end{acknowledgments}

\bibliography{main}

\end{document}